\documentstyle[12pt]{article}
\newcommand\bea{\begin{eqnarray}}
\newcommand\eea{\end{eqnarray}}

\newcommand\A{\alpha_z}
\newcommand\g{\gamma}
\newcommand\G{\Gamma}
\newcommand\m{M_l^m}
\newcommand\y{Y_l^m}
\newcommand\T{\theta}
\newcommand\f{\phi}
\newcommand\D{\delta}

\begin{document}
\thispagestyle{empty}
\bibliographystyle{unsrt}
\setlength{\baselineskip}{18pt}
\parindent 24pt
%\vspace{60pt}

%\begin{titlepage}
%\title
\begin{center}{
{\Large \bf
Multipole decomposition of potentials\\
in relativistic heavy ion collisions
}\\
%\vskip 1truecm
%\author{
A. Isar\\
{\it Department of Theoretical Physics, Institute of Atomic Physics\\
POB MG-6, Bucharest-Magurele, Romania\\
Internet: {\rm isar@theory.nipne.ro}}\\

}\end{center}
%\maketitle
%\vskip 1truecm
\begin{abstract}
%\vskip 0.5truecm

In relativistic heavy ion collisions an exact multipole decomposition of the
Lorentz transformed time
dependent Coulomb potentials
in a coordinate system with equal constant, but opposite
velocities of the ions,
is obtained for both
zero and different from zero impact parameter. The case of large
values of $\g$ and the gauge transformation of the interaction
removing both the $\g$ dependence and the $\ln b$ dependence are
also considered.
\end{abstract}
%\vskip 3truecm
%Shortened title:

%PACS numbers:
%\end{titlepage}
%\newpage
%\setcounter{page}{1}
\section{Introduction}

Relativistic heavy ion collisions offer us the opportunity to
investigate the properties of nuclear matter under very extreme
conditions of energy and temperature \cite{sto}. Information on
the collision processes is only obtained indirectly from the
energy and angular distributions of the produced particles and
from fragment distributions. Electromagnetic probes like the
photon and dilepton can yield esential information on the
evolution of relativistic heavy ion collisions, since these
signals leave the reaction zone without distorsion by strong
interaction. Photon and dilepton production are traditionally
considered to be one of the possible probes for detecting the
quark-gluon plasma \cite{shu}.

Relativistic heavy ion collisions provide a tool for the
investigation of electrons in extremely strong electromagnetic
fields  \cite{rsof}. Different methods were used for studying
the atomic physics effects -- coupled channel equations,
perturbation theory \cite{eic}, finite difference equations
(electron-positron pair production, pair production with capture).
The field of electron-positron pair creation in relativistic
heavy ion collisions is reviwed in Refs. \cite{eic,ber}. Different
processes were studied in Refs. [6-9].
%\cite{mom,rmom,thi,mgru}.
Electron-positron pair creation in relativistic
heavy ion collisions was non-perturbatively described by coupled
channel equations in a coordinate system with equal constant,
but opposite velocities of the ions \cite{thof}. Electron-positron pairs are
produced by the electromagnetic fields of high $Z$-ions in
atomic collisions at relativistic velocities. Atomic collisions
mean collisions with impact parameters which are larger than the
sum of the nuclear radii, i.e., the nuclear forces have no
influence on the collision.
The problem of dileptons excited
through the mechanism of virtual bremsstrahlung was investigated
in Ref. \cite{wei}.

In Sec. II we examine the basic step of the calculational
technology: the multipole expansion. An expansion in $1/\gamma$
produces the multipole expansion as compact, explicit and simple
functions of space and time. We will fix the coordinate system
in the centrum of mass of the colliding ions, which is the more
appropriate system for studying the multiple pair production
process. We proceed to a multipole decomposition of the two
potentials, in order to approximate the two-center Dirac
equation by a simpler one-center Dirac equation. We shall
consider both the cases of large values of $\g$ and exact
multipole decomposition, for both different from zero and zero
impact parameter. In Sec. III we discuss about the gauge
transformation of the interaction removing both the $\g$
dependence and the $\ln b$ dependence.
%and in Sec. IV we give a summary.

\section{Classical form of the interaction}

In the consideration of the electromagnetic interaction of
relativistic heavy ions, their motion is assumed to be well
described by straight line, classical paths and unperturbed by
recoil effects. Since we are interested in the multiple pair
production process, it is more natural to fix the coordinate
system in the centrum of mass (or center of momentum -- we
consider the same type of colliding ions, with the same $A$ and
$Z$ and equal constant, but opposite velocities) of the colliding ions, rather
than
on one of the ions, which is the case, for example, for bound
electron-positron creation. The interaction of the
electron-positron field with the moving ions is then given by
the Lorentz transformed Coulomb potentials (Li\' enard-Wiechert potentials)
\bea
V_1(\mbox{\boldmath $\rho$},z,t)={eZ(1-v{\bf \alpha})\over \{
[({\bf b}/2-\mbox{\boldmath $\rho$})/\g]^2
+(z-vt)^2\}^{1/2}},
\eea
\bea
V_2(\mbox{\boldmath $\rho$},z,t)={eZ(1+v\A)\over \{[(-{\bf b}/2-\mbox
{\boldmath $\rho$})/\g]^2+
(z+vt)^2\}^{1/2}}.
\eea
Here $b$ is the distance between the ion straight line paths,
which are taken along the $z$ axis (see Fig. 1); $eZ$ are the charges and
${\bf v}, -{\bf v}$ the velocities of the ions, $\g=1/(1-v^2)^{1/2}$,
$\mbox{\boldmath $\rho$}, z$ and $t$ are the coordinates of the
electron-positron
field relative to the c.m. system and $\A$ is the Dirac matrix.
The characteristic effects of the electromagnetic interactions
$eV_1$ and $eV_2$ produced by the ions are contained in their time
dependence and the very severe compression of the spatial
dependences (equivalently -- the sharp pulse description).
The time-dependent two-center Dirac equation for the
electron-positron motion is given by
\bea
i{\partial\over\partial t}\Psi({\bf r}, t)=H_D\Psi({\bf r},t)
\eea
(we employ the natural unit system with $\hbar=c=m_e=1).$
The Dirac Hamiltonian $H_D$ is given by
\bea
H_D=H_0-eV_1-eV_2,
\eea
\bea
H_0=\mbox{\boldmath $\alpha$}\cdot{\bf p}+\beta.
\eea

Since a method for constructing continuum wave functions for the
two-center Dirac equation is known only for the static case
(only for scalar potentials) \cite{wie,rum}, we proceed to a
multipole decomposition of $V_1$ and $V_2,$ in order to
approximate the two-center Dirac equation (3) by a simpler
one-center Dirac equation. We will consider the following cases:

A. - large values of $\g$ and $b\not= 0$ (but not too large);

B. - exact multipole decomposition and $b\not= 0$;

C. - exact multipole decomposition and $b=0$;

D. - large values of $\g$ and $b=0.$

\subsection {A. The case of large values of $\g$ and $b\not= 0$
(but not too large)}

In \cite{bal1} there were derived simple closed forms for the
multipole $\m$ that are accurate for large values of $\g$
(asymptotic-large $\gamma$ limit) for
the transformed Coulomb potential $V_p$ in the case of a moving
projectile and the coordinate system fixed on the target nucleus:
\bea
V_p(\mbox{\boldmath $\rho$},z,t)={eZ_p(1-v_p\A)\over \{[({\bf b}-\mbox
{\boldmath $\rho$})/\g]^2
+(z-v_p t)^2\}^{1/2}},\eea
where $Z_p,v_p$ and ${\bf b}$ are respectively the charge, the
velocity and the impact parameter of the projectile. This
multipole decomposition of the potential $V_p$ reads
\bea
\m(r,t)=\int d\Omega\y{1\over \{[({\bf b}-\mbox{\boldmath $\rho$})/\g]^2+(z-v_p
t)^2\}^{1/2}}, \eea
\bea
V_p(\mbox{\boldmath $\rho$},z,t)=e
Z_p(1-v_p\A)\sum_{l,m}\m(r,t)Y_l^{m*}(\T,\f)
=eZ_p(1-v_p\A)\sum_{l,m}V_l^m({\bf r},t).
\eea
The derivation consists of an almost straightforward expansion
about the near singular point $z=v_p t,$ after some careful
rearrangements of the integrand.
It is important to note that the expansion requires that ${b/
(r-v_p t)}\ll \g;$ assuming that the nature of the physical
problem confines $r$ and $v_p t$ to moderate multiples of
$\hbar/ m_e c,$ the requirement of $b$ becomes $b \ll \g\hbar/
m_e c$ \cite{bal1}.
Defining $\T_t$ by
\bea
(1)~~ |v_p t|< r,~~ \cos\T_t={v_p t\over r},~~ \sin\T_t=
\sqrt{1-({v_pt\over r})^2}, \nonumber \\
(2)~~ |v_p t|> r,~~ \cos \T_t=1,~~ \sin \T_t=0,
\eea
replacing $v_p$ by 1 (they differ only by $O({1/\g^2})$ and
noting that to order $\ln\g/\g^2$ the $({\bf b}-\mbox{\boldmath $\rho$})^2
/\g^2$ term
in the denominator can be replaced by its value at $\T=\T_t,$ the
asymptotic form of the multipole decomposition of $V_p(\mbox{\boldmath
$\rho$},z,t)$ (where $z=r\cos\T$)
is given for $m>0$ by \cite{bal1,bal2}
\bea
\m(r,t)={\y({\displaystyle t\over r},0)\over r}\times
\left\{\begin{array}{rcl}
 0, & r< t \\
{\displaystyle
{2\pi\over m}({r^2-t^2\over b^2})^{m\over 2}},
& t<r<\sqrt{b^2+t^2} \\
{\displaystyle {2\pi\over m}({b^2\over r^2-t^2})^{m\over 2}}, &
r>\sqrt{b^2+t^2}.
\end{array} \right.
\eea
This expression is valid for positive $m$ and $t.$ Negative $m$
and $t$ expressions are given by symmetry.
For $m=0,$ the
asymptotic form is \cite{bal1,bal2}
\bea
M_l^0(r,t)=
{\sqrt{\pi(2l+1)}\over r}
\times
\left\{\begin{array}{rcl}
  2Q_l({\displaystyle {t\over r}}), & r< t \\
P_l({\displaystyle {t\over r}})(2\ln 2\g +\ln
{\displaystyle {r^2-t^2\over b^2}}
-2{\displaystyle \sum_{n=1}^l{1\over n}}), & t<r<\sqrt{b^2+t^2} \\
P_l({\displaystyle {t\over r}})(2\ln 2\g -2{\displaystyle \sum_{n=1}^l
{1\over n}}), & r>\sqrt{b^2+t^2},
\end{array} \right.
\eea
valid for positive $t$, with the negative $t$ form again to be
obtained by symmetry. Here $Q_l$ is the Legendre function of the second kind.
Unless the $m=0$ multipole terms are to be integrated together
with functions that vary sharply in the $r\sim v_p t$ region,
the need for the bridge function at $r\sim t$ disappears \cite{bal1}.
The evaluation of time-dependent matrix elements for use in
coupled-channels calculations will be facilitated by the form of
the above interaction (which will have an analogous structure
for our two-center case). In Eqs. (10),(11) all terms can be expressed
in terms of polynomials and/or convergent series in $t/r$
for $t<r$ and in $r/t$ for $r<t.$ Thus the matrix
element to be integrated over $r$ can have its interaction
expressed as a series of negative and positive powers of $r.$ In
each of the terms of the series in $r,$ $t$ is effectivelly a
coefficient: $r$ and $t$ have become separable term by term.
These asymptotic forms provide a simple means of calculating
matrix elements. In addition they allow to straightforwardly
look at the $\g$ dependence of any set of matrix elements or
amplitudes. As one can see directly, the $\g$ dependence of the
multipole operators appears only in the $m=0$ terms and only in
the one form \cite{bal1}
\bea
2(\ln 2\g)eZ_p(1-\A)\sum_lY_l^0(\T){\sqrt{\pi(2l+1)}\over
r}P_l({t\over r})=2(\ln 2\g)eZ_p(1-\A)\delta(z-t), ~r>t;
\eea
the condition $r\ge t$ is automatically contained in the
$\delta(z-t).$ But as we shall see in Sec. III, this set of
terms is entirely removable by a gauge transformation. Once
having removed this explicit dependence on $\g,$ we are left
only with the implicit dependence inherent in the limitation
already noted that the above forms are valid only for the region
$b\ll\g\hbar/m_e c.$ However, when $b$ is not small,
perturbation theory is valid.

The multipole expansion for the potential $V_1$ is given also by
the formulas (10),(11), but with $b$ replaced by $b/2, Z_p$ by $Z$ and
$v_p$ by $v$ with $v=1.$ We have proven that the multipole
expansion for $V_2$ can be obtained from that for $V_1,$
by replacing $1-\A$ by $1+\A$ and by multiplying the whole expression
by $(-1)^l.$

1. {\it Monopole approximation}

In the case $l=0,m=0,$ we have
\bea
V_1^{(0)}=eZ(1-\A){1\over 2r} \times
\left\{\begin{array}{rcl}
 2Q_0({\displaystyle {t\over r}}), & r< t \\
2\ln 2\g +{\displaystyle \ln{r^2-t^2\over
b^2/4}}, & t< r<
\sqrt{b^2/4+t^2} \\
2\ln 2\g,& r>
\sqrt{b^2/4+t^2}
\end{array} \right.
\eea
and $V_2^{(0)}=V_1^{(0)}(\A\to - \A).$
Then
\bea
V^{(0)}=V_1^{(0)}+V_2^{(0)}=
{2eZ\over r}\times
\left\{\begin{array}{rcl}
 Q_0({\displaystyle {t\over r}}), & r< t \\
\ln 2\g +{\displaystyle{1\over 2}}\ln{\displaystyle {r^2-t^2\over
b^2/4}}, & t< r<
\sqrt{b^2/4+t^2} \\
\ln 2\g, & r> \sqrt{b^2/4+t^2}.
\end{array}\right.
\eea
Here
\bea
Q_0({t\over r})={1\over 2}\ln{t+r\over t-r}.
\eea

2. {\it Dipole approximation ($l=1$)}

The dipole term is given by
\bea
V^{(1)}=V_1^{(1)}+V_2^{(1)}=
-{3eZ\A\over r}[
\sin\T\cos\f\sin\T_t\times
\left\{\begin{array}{rcl}
0, & r< t \\
{\displaystyle({r^2-t^2\over b^2/4})^{1\over 2}}, & t< r< \sqrt{b^2/4+t^2} \\
{\displaystyle({b^2/4\over r^2-t^2})^{1\over 2}}, & r> \sqrt{b^2/4+t^2}
\end{array} \right. \nonumber
\eea
\bea
+\cos\T \times
\left\{\begin{array}{rcl}
2 Q_1({\displaystyle {t\over r}}), & r< t \\
{\displaystyle
{t\over r}}(2\ln 2\g +\ln{\displaystyle {r^2-t^2\over
b^2/4}}-2), & t<r<\sqrt{b^2/4+t^2} \\
2{\displaystyle {t\over r}}(\ln 2\g-1), & r>\sqrt{b^2/4+t^2}.
\end{array}\right.
].
\eea
Here
\bea
Q_1({t\over r})={t\over r}Q_0({t\over r})-1={t\over 2r}\ln{t+r\over t-r}-1.
\eea
We remind that $v=1$ in the case A.
In the dipole approximation we have to add the terms $V^{(0)}$
and $V^{(1)}$ given by Eqs. (14) and (16):
\bea
V=V^{(0)}+V^{(1)}.
\eea

\subsection {B. Exact multipole decomposition ($b\not=0$)}

The exact multipole decomposition is given as the series \cite{bal1}:
\bea
V_1=eZ(1-v\A)\sum_{l,m}\y(\T,\f)\sum_{l'}Y_{l'}^m(\T_u,0){\cal R}(l,l';r,u)
{\cal A}(m;l,l';{1\over v}),
\eea
where
\bea
u=[{b^2\over 4}+ (vt)^2]^{1/2},~~ \cos\T_u={vt\over u},
~~\sin\T_u={b\over 2u}
\eea
and
\bea
{\cal R}(l,l';r,u)={\pi\over 4}{u^{l'}\over
r^{l'+1}}{\G({l+l'+1\over 2})\over\G({l-l'\over
2}+1)\G(l'+{3\over 2})}F({l+l'+1\over2},{l'-l\over
2},{l'+{3\over 2},{u^2\over r^2}})~~ {\rm for}~~ r>u,
\eea
\bea
{\cal A}(m;l,l';{1\over v})={1\over 4\pi}(-1)^m\sqrt{(2l+1)(2l'+1)}
\sum_LC_{-mm0}^{ll'L}C_{000}^{ll'L}{2\over v}Q_L({1\over v}).
\eea
For $r<u,$ we have to do the interchange $r\leftrightarrow u,$
$l\leftrightarrow l'$ in ${\cal R}.$ For the special case $r=u,$
\bea
{\cal R}(l,l';r,r)={\pi\over 4}{1\over
r}
{\displaystyle
{\G({l+l'+1\over 2})\over\G({l-l'\over
2}+1)\G({l+l'\over 2}+1)\G({l-l'\over 2}+1)}.
}
\eea
This form for the multipole expansion is equivalent with that
appearing in the literature. There is a set of special cases
that is particularly simple \cite{bal1}. For $r>u$ the sum over
$l'$ in Eq. (19) is cut off by the spherical Bessel function integral (for
${\cal R})$ (the occurence of a negative-integer value of the
argument of the gamma function ensures that $l'\le l);$
furthermore, for $l'\le l$ the hypergeometric function is just
a polynomial.
Thus the exact result for $Y_1^1$ component of $V_1$ is
for $r>u$:
\bea
Y_1^1(\T,\f)\sqrt{{3\pi\over 2}}{b\over 2r^2}\left[{1\over v}({1\over
v^2}-1)Q_0({1\over v})-{1\over v^2}\right].
\eea
As $\g\to\infty, v\to 1,$ then the term in square brackets
approaches $-1,$ in agreement with the corresponding expression in
the case A. Similarly, the $Y_1^0$ component of $V_1$ is for $r>u$:
\bea
Y_1^0(\T,\f)\sqrt{3\pi}{t\over r}\left[{2\over v^2}
Q_0({1\over v})-2\right]. \eea
As $\g\to\infty,$ the term in square brackets
approaches $2(\ln 2\g-1),$ in agreement with the case A.
The difference between exact and asymptotic values is $O(\ln\g/\g^2).$

Another form of the exact multipole decomposition can be
obtained if we write the denominator of $V_1$ as
\bea {1\over \sqrt{(x-b/2)^2+y^2+\g^2(z-vt)^2}}={1\over
|{\bf r}'-{\bf R}'(t)|},
\eea
with the vectors ${\bf r}'=(x,y,\g z)$ and ${\bf
R}'=({b/2}, 0, \g vt).$
The expansion in multipoles is then given by \cite{rsof,meh}:
\bea
{1\over |{\bf r}'-{\bf R}'|}=
\left\{\begin{array}{rcl}
{\displaystyle
\sum_{l=0}^\infty{4\pi\over 2l+1}{r'^l\over
R'^{l+1}}\sum_{m=-l}^lY_l^{m*}(\hat R')\y(\hat r'),
}
& r'\le R' \\
{\displaystyle
\sum_{l=0}^\infty{4\pi\over 2l+1}{R'^l\over
r'^{l+1}}\sum_{m=-l}^lY_l^{m*}(\hat R')\y(\hat r'),
}
& r'\ge R'.
\end{array}\right.
\eea
$\hat R'$ and $\hat r'$ represent the angular arguments of the
spherical harmonics $\y$  which give the direction of ${\bf R}'$
and ${\bf r}'.$
As usual the scattering plane is taken to be $\f=0.$ In
consequence the spherical harmonics $Y_l^{m*}(\hat R')$ is real-valued.

{\it 1. Monopole approximation $(l=0)$}

We have (with $\g^2v^2=\g^2-1$ and
$R'=\sqrt{{b^2/4}+\g^2v^2t^2}/\g$):
\bea
V_1^{(0)}=eZ(1-v\A)\times
\left\{\begin{array}{rcl}
{\displaystyle
{1\over R'},
} & r'\le R' \\
{\displaystyle{1\over r'}}, & r'\ge R'
\end{array}\right.
\eea
and $V_2^{(0)}=V_1^{(0)}(\A\to - \A).$
Then
\bea
V^{(0)}=2eZ\times
\left\{\begin{array}{rcl}
{\displaystyle {1\over R'}}, & r'\le R'\\
{\displaystyle {1\over r'}}, & r'\ge R.'
\end{array}\right.
\eea

{\it 2. Dipole approximation}

For $(l=1)$ we have:
\bea
V_1^{(1)}=eZ(1-v\A)\times
\left\{\begin{array}{rcl}
{\displaystyle
{r'\over R'^2}}\cos\D, & r'\le R'\\
{\displaystyle
{R'\over r'^2}}\cos\D, & r'\ge R',
\end{array}\right.
\eea
where $\D$ is the angle between the vectors ${\bf r}'$ and ${\bf R}'$
and $V_2^{(1)}=-V_1^{(1)}(\A\to - \A).$
Then the dipole term is given by
\bea
V^{(1)}=-2eZv\A\cos\D\times
\left\{\begin{array}{rcl}
{\displaystyle{r'\over R'^2}}, & r'\le R'\\
{\displaystyle{R'\over r'^2}}, & r'\ge R'
\end{array}\right.
\eea
and in the dipole approximation
\bea
V=V^{(0)}+V^{(1)}=2eZ
\left[\times\left\{\begin{array}{rcl}
{\displaystyle
{1\over R'}} \\
{\displaystyle
{1\over r'}}
\end{array}\right.
-v\A\cos\D\times
\left\{\begin{array}{rcl}
{\displaystyle{r'\over R'^2}}, & r'\le R'\\
{\displaystyle{R'\over r'^2}}, & r'\ge R'
\end{array}\right.
\right].
\eea

\subsection{C. Exact multipole decomposition. The case $b=0$}

For the calculation of electron-positron pair production in
heavy ion reactions the situation as $b\to 0$ is of particular
interest because it is in this lowest impact regime that
the failure of perturbation theory is greatest. Since the scale at which
nuclear effects contribute for two interacting large nuclei ($b\le
15$ fm) is much smaller than the atomic scale for electrons and
positrons ($\hbar/m_e c=386$ fm), the $b=0$ solution should be of
interest as closely approximating the situation at the
smallest impact parameter for which nuclear effects are not
applicable (or for which we omit all nuclear reactions).
Sample calculations show $b=0$ and $b=15$ fm results to be $\le
1\%$ apart \cite{bal2,bal3}.

Let us consider the $b=0$ case without any approximation in
$1/\g.$ Here we assume $t>0.$ To obtain appropriate expressions
for $t<0$ one can make use of the symmetry relation
$M_l^0(r,-t)=(-1)^lM_l^0(r,t).$
By definition we have \cite{bal2,bal3} in the case of $V_1:$
\bea
M_l^{m(1)}(r,t)=\int d\Omega \y{1\over \{\rho^2/\g^2+(z-vt)^2\}^{1/2}}.
\eea
By symmetry only $m=0$ is non-vanishing and we immediately
integrate over $\f$$(\rho=r\sin\T, z=r\cos\T):$
\bea
M_l^{m(1)}(r,t)=2\pi\D_{m,0}\int_{-1}^1 {Y_l^0(\cos\T)d(\cos\T)
\over \{r^2\sin^2\T/\g^2+(r\cos\T-vt)^2\}^{1/2}}.
\eea
Let us define $x=\cos\T$ and note that $1/\g^2=1-v^2$ to rewrite
\bea
M_l^{m(1)}(r,t)={\sqrt{2\pi(2l+1)}\over rv}\int_{-1}^1
{P_l(x) dx
\over
{\displaystyle
({1\over v^2} +{t^2\over r^2}-1-{2xt\over vr}+x^2)^{1/2}}
}.
\eea
This expression is symmetric in $1/v$ and $t/r$
and the integral over $x$ can be carried out in closed form for
each value of $l.$ The authors of Refs. \cite{bal2,bal3} recently found the
following simple exact result for the multipole moments of the
$b=0$ case:
\bea
M_l^{m(1)}(r,t)=\D_{m,0}{2\sqrt{\pi(2l+1)}\over rv}\times
\left\{\begin{array}{rcl}
P_l({\displaystyle {1\over v}})Q_l({\displaystyle {t\over r}}), & r<vt \\
{\displaystyle Q_l({1\over v})P_l({t\over r})}, & r>vt.
\end{array}\right.
\eea
Then
\bea
V_1=eZ(1-v\A)\sum_lY_l^0(\T){2\sqrt{\pi(2l+1)}\over rv}\times
\left\{\begin{array}{rcl}
{\displaystyle
P_l({1\over v})Q_l({t\over r})}, & r<vt \\
{\displaystyle Q_l({1\over v})P_l({t\over r})}, & r>vt.
\end{array}\right.
\eea
We obtain for $V_2$ the expression
\bea
V_2=eZ(1+v\A)\sum_l(-1)^lY_l^0(\T){2\sqrt{\pi(2l+1)}\over rv}\times
\left\{\begin{array}{rcl}
{\displaystyle
P_l({1\over v})Q_l({t\over r})}, & r<vt \\
{\displaystyle Q_l({1\over v})P_l({t\over r})}, & r>vt.
\end{array}\right.
\eea

{\it 1. Monopole approximation $(l=0)$}

We obtain:
\bea
V^{(0)}=V_1^{(0)}+V_2^{(0)}=
{2eZ\over r} {1\over v}\times
\left\{\begin{array}{rcl}
 Q_0({\displaystyle {t\over r}}), & r<vt \\
 Q_0({\displaystyle {1\over v}}), & r>vt.
\end{array}\right.
\eea
By taking the limit $v\to 0,$${\displaystyle
\lim_{v\to 0} {1\over v}Q_0({1\over v})=1
},$
so that in the non-relativistic limit we have
$V_{non-rel}=2eZ/r,$ as it should be.

{\it 2. Dipole approximation $(l=1)$}

In this case we have:
\bea
V_1^{(1)}=
eZ(1-v\A)Y_1^0(\T){2\sqrt{3\pi}\over rv}\times
\left\{\begin{array}{rcl}
{\displaystyle
{1\over v}Q_1({t\over r})}, & r< vt \\
{\displaystyle {t\over r}Q_1({1\over v})}, & r> vt
\end{array}\right.
\eea
and $V_2^{(1)}=-V_1^{(1)}(\A\to - \A).$
Then
\bea
V^{(1)}=V_1^{(1)}+V_2^{(1)}=
-{6eZ\A\over r}\cos\T \times
\left\{\begin{array}{rcl}
{\displaystyle{1\over v}Q_1({t\over r})}, & r< vt \\
{\displaystyle{t\over r}Q_1({1\over v})}, & r> vt.
\end{array}\right.
\eea
Therefore in the dipole approximation
\bea
V=V^{(0)}+V^{(1)}=
{2eZ\over r} \left[{1\over v}\times
\left\{\begin{array}{rcl}
{\displaystyle
 Q_0({t\over r})} \\
{\displaystyle Q_0({1\over v})
}
\end{array}\right.
-3\A \cos \T \times
\left\{\begin{array}{rcl}
{\displaystyle
{1\over v}Q_1({t\over r})}, & r< vt \\
{\displaystyle{t\over r}Q_1({1\over v})},  & r> vt
\end{array}\right.
\right].
\eea

\subsection{D. The case of large values of $\g$ and $b=0$}

For $b=0$ the expression of multipole moments takes on the
simple form good up to order $\ln\g/\g^2$ \cite{bal2,bal3}.
For $V_1:$
\bea
M_l^{m(1)}(r,t)=\D_{m,0}
{2\sqrt{\pi(2l+1)}\over r}\times
\left\{\begin{array}{rcl}
{\displaystyle
  Q_l({t\over r})}, & r<t  \\
{\displaystyle P_l({t\over r})}(\ln 2\g -{\displaystyle
\sum_{n=1}^l{1\over n}}), & r>t.
\end{array}\right.
\eea
Here $t>0.$ To obtain appropriate expressions for $t<0$ one can
make use again of the symmetry relation.
Analogously, we have for $V_2:$
$M_l^{m(2)}(r,t)=(-1)^lM_l^{m(1)}(r,t).$
In this asymptotic form, the split between the two regions
$r>vt,r<vt$ has been written as $r>t,r<t,$ dropping the $1/\g^2$
differences (we remind that $v\to 1).$

{\it 1. Monopole approximation $(l=0)$}

We have:
\bea
V_1^{(0)}={eZ(1-\A)\over r}\times
\left\{\begin{array}{rcl}
Q_0({\displaystyle{t\over r}}), & r< t \\
\ln 2\g, & r>t
\end{array}\right.
\eea
and $V_2^{(0)}=V_1^{(0)}(\A\to - \A).$
Then
\bea
V^{(0)}=V_1^{(0)}+V_2^{(0)}=
{2eZ\over r}\times
\left\{\begin{array}{rcl}
  Q_0({\displaystyle{t\over r}}), & r< t \\
\ln 2\g,  & r>t.
\end{array}\right.
\eea

{\it 2. Dipole approximation $(l=1)$}

We have:
\bea
V_1^{(1)}=
eZ(1-\A)Y_1^0(\T){2\sqrt{3\pi}\over r}\times
\left\{\begin{array}{rcl}
Q_1({\displaystyle {t\over r}}), & r< t \\
{\displaystyle {t\over r}}(\ln 2\g-1), & r> t
\end{array}\right.
\eea
and $V_2^{(1)}=-V_1^{(1)}(\A\to - \A).$
Then
\bea
V^{(1)}=-{6eZ\A\over r}\cos\T\times
\left\{\begin{array}{rcl}
Q_1({\displaystyle {t\over r}}), & r< t \\
{\displaystyle {t\over r}}(\ln 2\g-1), & r> t.
\end{array}\right.
\eea
Therefore in the dipole approximation
\bea
V=V^{(0)}+V^{(1)}=
{2eZ\over r}\left[\times
\left\{\begin{array}{rcl}
{\displaystyle Q_0({t\over r})} \\
\ln 2\g
\end{array}\right.
-3 \A \cos\T\times
\left\{\begin{array}{rcl}
{\displaystyle
Q_1({t\over r})}, & r< t \\
{\displaystyle{t\over r}}(\ln 2\g-1), & r> t
\end{array}\right.
\right].
\eea

In the monopole approximation, where only the term with $l=0$ in
the multipole expansion of the two-center potential is taken
into account, a charged spherical shell of radius $vt$ (case C)
or $t$ (case D) simulates the Coulomb potential of the two nuclei.

\section{Gauge transformation of the interaction}

It has previously been shown that
in the large $\gamma$ limit
a gauge transformation on the considered potential can remove both the large
positive and negative time contributions of the potential as well as the
$\gamma$ dependence \cite{bal2,bal3}.
If one makes the gauge
transformation on the wave function \cite{bal2,bal3}
\begin{equation}
\psi=e^{-i\chi({\bf r},t)} \psi',
\end{equation}
where
\begin{equation}
\chi({\bf r},t)={eZ \over v} \ln [\gamma(z-v t)+\sqrt{b^2+\gamma^2(z-v
t)^2}],
\end{equation}
the interaction $V(\mbox{\boldmath $\rho$},z,t)$ is gauge transformed to
\begin{equation}
V(\mbox{\boldmath $ \rho$},z,t)={eZ(1-v\alpha_z)\over
\sqrt{ [({\bf b}-\mbox{\boldmath $ \rho$})/\gamma]^2+(z-v t)^2}}-{e
Z(1-(1/v)\alpha_z)\over\sqrt{b^2/\gamma^2+(z-v t)^2}}.
\end{equation}
This transformation  removes both the $\g$ dependence and the
$\ln b$ dependence of the interaction for the region $r<\sqrt{b^2+t^2}.$
The large $\g$ expression for the $m=0$ gauge transformed
interaction then becomes \cite{bal2,bal3}
\bea
M_l^0(r,t)=
{\sqrt{\pi(2l+1)}\over r} P_l({t\over r})\times
\left\{\begin{array}{rcl}
0, & r<\sqrt{b^2+t^2}  \\
-\ln {\displaystyle{r^2-t^2\over b^2}}, & r>\sqrt{b^2+t^2}.
\end{array}\right.
\eea
Note that this transformation affects the $m=0$ part of the
interaction only.
An additional computational advantage of the previous form Eq.
(52) is
that the large positive and negative time contributions inherent
in the term $Q_l(t/r)$ of Eq. (11) have been removed, as was
shown by Baltz, Rhoades-Brown and Weneser \cite{bal2}.
It is very important that there is no $\g$ dependence in either
$m=0$ or $m\neq 0$ parts of the interaction.

In the untransformed interaction, Eq. (11), the large negative
and positive time interaction is dominated by the $m=0$
interaction for $t>r,$
\bea
V_l^0({\bf r},t)=Y_l^0(\T)2\sqrt{\pi}{\sqrt{2l+1}\over r}Q_l(t/r)\eea
(recall that there is no $t>r$ interaction for $m\neq 0$).
There is a similar $Q_l(t/r)$ dependence in the exact $b=0$
expression \cite{bal2}.
For the monopole interaction
\bea
V_0^0(r,t)={1\over 2r}\ln{1+r/t\over 1-r/t}={1\over t}+{r^2\over
3t^3}+{r^4\over 5t^5}+...\eea
By analyzing $Q_l(t/r)$ for $l\neq 0,$ one finds that other
$m=0$ interactions, $V_l^0(r,t)$ have terms of longest range
going as ${\displaystyle{1\over t^{l+1}}}.$
For $r<t$ we must consider $Q_l(t/r)$ which has a piece equal to
$-\ln(t-r).$
Likewise, for $t<r<\sqrt{b^2+t^2},$ we must consider \cite{bal2}
\bea
\ln{r^2-t^2\over b^2}=\ln{r+t\over b^2}+\ln(r-t).\eea

1. {\it Monopole approximation}

In the case $l=0,m=0,$ we have
\bea
V_1^{(0)}=eZ(1-\A){1\over 2r} \times
\left\{\begin{array}{rcl}
0, & r<\sqrt{b^2/4+t^2}  \\
-\ln {\displaystyle{r^2-t^2\over b^2/4}}, & r>\sqrt{b^2/4+t^2}
\end{array} \right.
\eea
and
$V_2^{(0)}=V_1^{(0)}(\A\to - \A).$
Then
\bea
V_g^{(0)}=V_1^{(0)}+V_2^{(0)}=
{2eZ\over r}\times
\left\{\begin{array}{rcl}
0, & r<\sqrt{b^2/4+t^2}  \\
-\ln {\displaystyle{r^2-t^2\over b^2/4}}, & r>\sqrt{b^2/4+t^2}.
\end{array}\right.
\eea

2. {\it Dipole approximation ($l=1$)}

The dipole term is given by
\bea
V_g^{(1)}=V_1^{(1)}+V_2^{(1)}=
-{3eZ\A\over r}[
\sin\T\cos\f\sin\T_t\times
\left\{\begin{array}{rcl}
0, & r< t \\
{\displaystyle({r^2-t^2\over b^2/4})^{1\over 2}}, & t< r< \sqrt{b^2/4+t^2} \\
{\displaystyle({b^2/4\over r^2-t^2})^{1\over 2}}, & r> \sqrt{b^2/4+t^2}
\end{array} \right. \nonumber
\eea
\bea
+\cos\T \times
\left\{\begin{array}{rcl}
0, & r<\sqrt{b^2/4+t^2}  \\
-\ln {\displaystyle{r^2-t^2\over b^2/4}}, & r>\sqrt{b^2/4+t^2}.
\end{array}\right.
].
\eea
In the dipole approximation we have to add the terms $V_g^{(0)}$
and $V_g^{(1)}$ given by Eqs. (57) and (58):
\bea
V_g=V_g^{(0)}+V_g^{(1)}.
\eea

{\it Acknowledgments.} I would like to express my
gratitude to Professors N. Gr\" un and W. Scheid for helpful
discussions and their warm hospitality at the Institute of
Theoretical Physics, University of Giessen (Germany), where part
of this work was done.

\end{document}